\documentclass{article}

\usepackage{arxiv}

\usepackage[utf8]{inputenc} 
\usepackage[T1]{fontenc}    
\usepackage{hyperref}       
\usepackage{url}            
\usepackage{booktabs}       
\usepackage{amsfonts}       
\usepackage{nicefrac}       
\usepackage{microtype}      
\usepackage{lipsum}
\usepackage{graphicx}
\graphicspath{ {./images/} }

\title{Anomalous Photoresponse of Heavily Doped GaAs/AlAs Superlattices
with Electric Domains
}

\author{
 I. V. Altukhov, S. E. Dizhur$^{*}$, M. S. Kagan$^{**}$, N. A. Khvalkovskiy, S. K. Paprotskiy \\
  V.A. Kotel’nikov Institute of Radio Engineering and Electronics \\
  Russian Ac. Sci. \\
  Moscow, Russia \\
   \And
   I. S. Vasil'evskii, A. N. Vinichenko \\
   National Research Nuclear University MEPhI \\
  Moscow, Russia \\
   \And
  $^{*}$\texttt{sdizhur@phystech.edu} \\
  $^{**}$\texttt{kagan@cplire.ru} \\
}

\begin{document}
\maketitle
\begin{abstract}
The strong effect of weak interband illumination on tunneling transport in doped GaAs/AlAs superlattices was found under conditions of electric domain formation. The photoresponse at voltages below the threshold one (before the domain formation) did not observe. The phenomenon is referred to strong carrier depletion inside the triangular high-field domain. The domain modes transformations under the illumination were also found. 
\end{abstract}


The electron transport in semiconductor superlattices (SLs) has been studied in detail in recent decades~\cite{wacker, bonilla}, mainly in connection with the predicted amplification of Bloch waves, which can make it possible to create tunable sources of THz radiation. The main hindrance to the Bloch wave amplification is the formation of electrical domains owing to the static negative differential conductivity (NDC) arising in SLs at resonant tunneling between confined states in neighboring quantum wells (Esaki-Tsu mechanism~\cite{esakitsu}). The domain formation in SLs eliminates the Bloch gain. Many works were devoted also to studies of domain regimes (see, e.g.,~\cite{bonilla} and references therein). Most of them were performed with weakly doped SLs at low temperatures under strong illumination~\cite{bonilla, sun,tom,kwok1,kwok2,klitzing,grahn}. The change in free carrier concentration under illumination gives rise to different domain regimes. It has been shown, in particular, that transition from traveling to static domain regime occurs at sufficiently high illumination intensity. Presented in this work are studies of the effect of rather weak interband illumination on a tunneling current in highly doped short-period GaAs/AlAs superlattices with the high-field domains. 

\begin{figure}[h!] 
    \centering
    \includegraphics[scale=0.3]{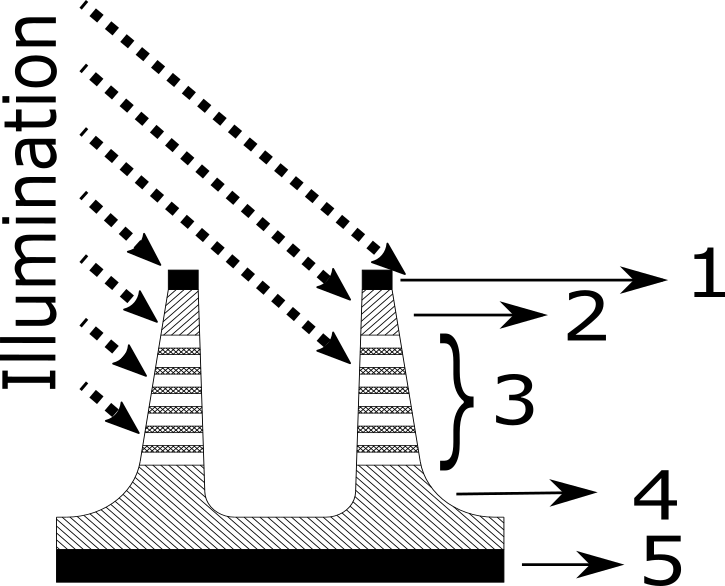}
    \caption{The sectional view of the structure. 1, 5 – top (ring) and bottom metal contacts, 2 – top contact n+ layer, 3 – superlattice, 4 – heavily doped substrate.}
    \label{fig:fig1}
\end{figure}

The GaAs/AlAs SLs grown by molecular beam epitaxy consisted of $100$ periods of $4$\, nm GaAs quantum wells (QWs) divided with $2$\, nm AlAs barriers. The SLs were sandwiched between heavily doped ($\sim 10^{19}$\, cm$^{-3}$) {\it n+} cap layer and {\it n+} substrate (Figure~\ref{fig:fig1}). 
GaAs quantum wells were doped with shallow donors (Si) at a concentration of $2 \times 10^{17}$\, cm$^{-3}$. The structures were processed with electron lithography techniques to produce ring-shaped mesa structures of $10$--$15$\, $\mu$m diameters and $0.8$--$1.5$\, $\mu$m widths. The samples were illuminated in the direction shown in Figure~\ref{fig:fig1}. Even at the interband illumination, the excited electron-hole pairs are distributed practically uniform on the ring width due to ambipolar diffusion which length is estimated as $\sim 5$\, $\mu$m. Triangular voltage pulses with sweep-up times of $0.2$ to $10$\, $\mu$s and sweep-down times up to $100$\, $\mu$s were applied to the samples. The time dependences of current and voltage were used to record current-voltage (I-V) curves at increasing and decreasing voltages. Measurements were performed at room temperature. 
	
\begin{figure}[h!] 
    \centering
    \includegraphics[scale=0.4]{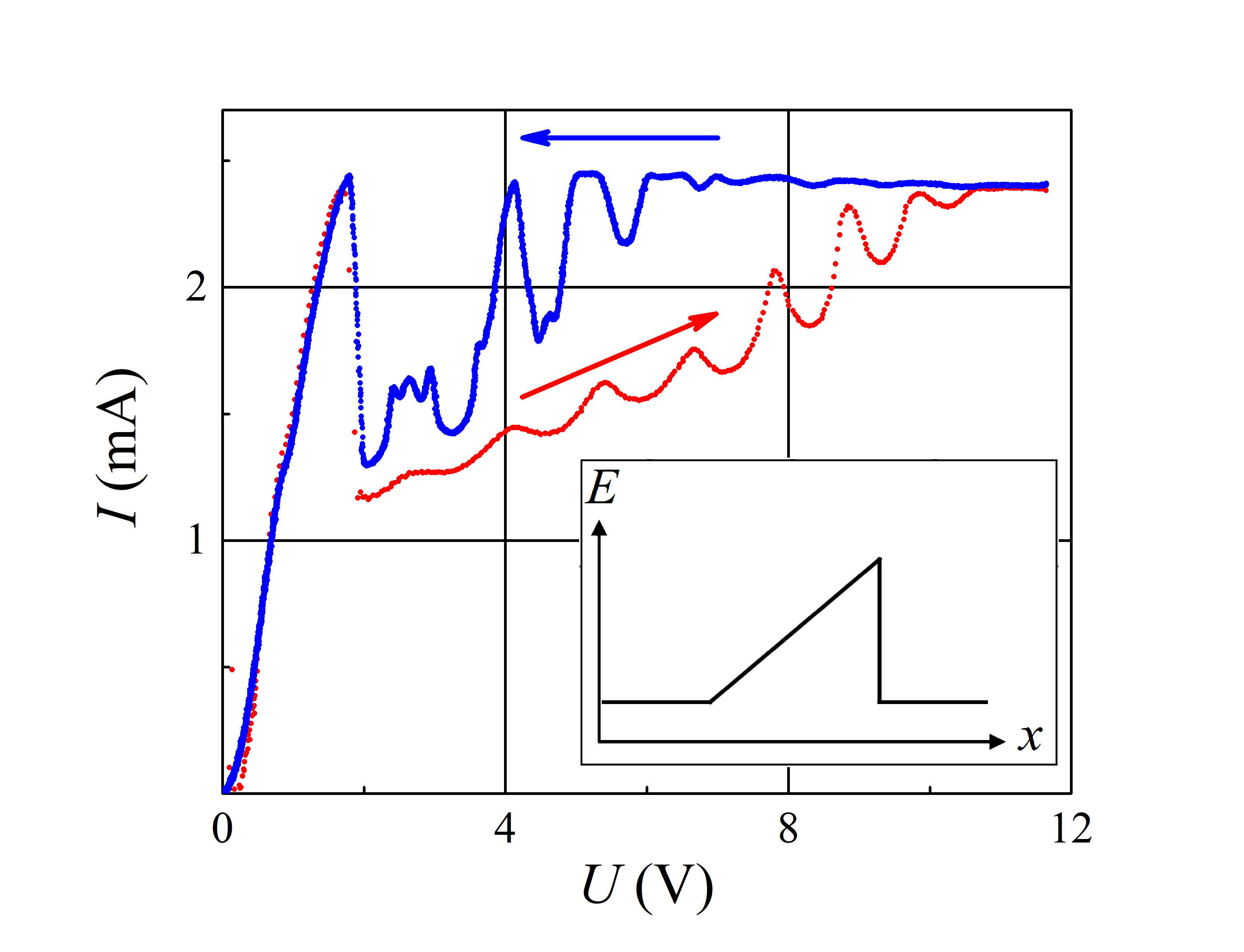}
    \caption{Current-voltage characteristics of the GaAs/AlAs SLs at sweep-up (red) and sweep-down (blue) of voltage. Arrows show the voltage sweep direction. Inset: field distribution inside the moving domain (schematically).}
    \label{fig:fig2}
\end{figure}

Shown in Figure~\ref{fig:fig2} are the I-V characteristics of GaAs/AlAs SLs at sweeping-up and sweeping-down voltage. 
A sharp drop in the current (up to $\sim 2$ times) at some threshold voltage $V_{th}$ indicates the formation of moving domains~\cite{wacker,bonilla}. The saturation regions in I-V curves are due to static domain formation. The current hysteresis can be explained by a transition between the traveling and the static high-field domain regimes~\cite{suris}. At the direct sweep, the saturation begins at the voltage $V > V_{th}$ when the current reaches a value at the maximum of the I–V characteristic. In the reverse scan, the saturation region expands to smaller fields. This is because the transition voltage from the moving to the static domain is larger than for the reverse transition with decreasing voltage. 

The current rise at voltages above the threshold gives evidence of the triangular shape of the dipole domain (see inset in Figure~\ref{fig:fig2}). The explanation of the domain shape is as follows~\cite{bonilla,suris}. A rough estimation of electric field values outside and inside the domain using the experimental I-V curves gives the carrier concentration $\sim 5 \times 10^{18}$\, cm$^{-3}$, which ensures the necessary field step at the domain boundaries (dipole domain ''walls''). A negative charge in one of the domain walls is created by free electrons, which can be accumulated in one QW. The positive charge at the opposite domain wall is due to positively charged donors, the concentration of which is $\sim 2 \times 10^{17}$\, cm$^{-3}$ being more than one order less. Because of the depopulation of donors, linear spatial growth of electric field inside the domain appears. The series of almost voltage-periodic maxima on a background of the current rise in I-V curves is attributed to the optical phonon-assisted tunneling between quantum wells inside the triangular domain~\cite{suris}. 

\begin{figure}[h!] 
    \centering
    \includegraphics[scale=0.4]{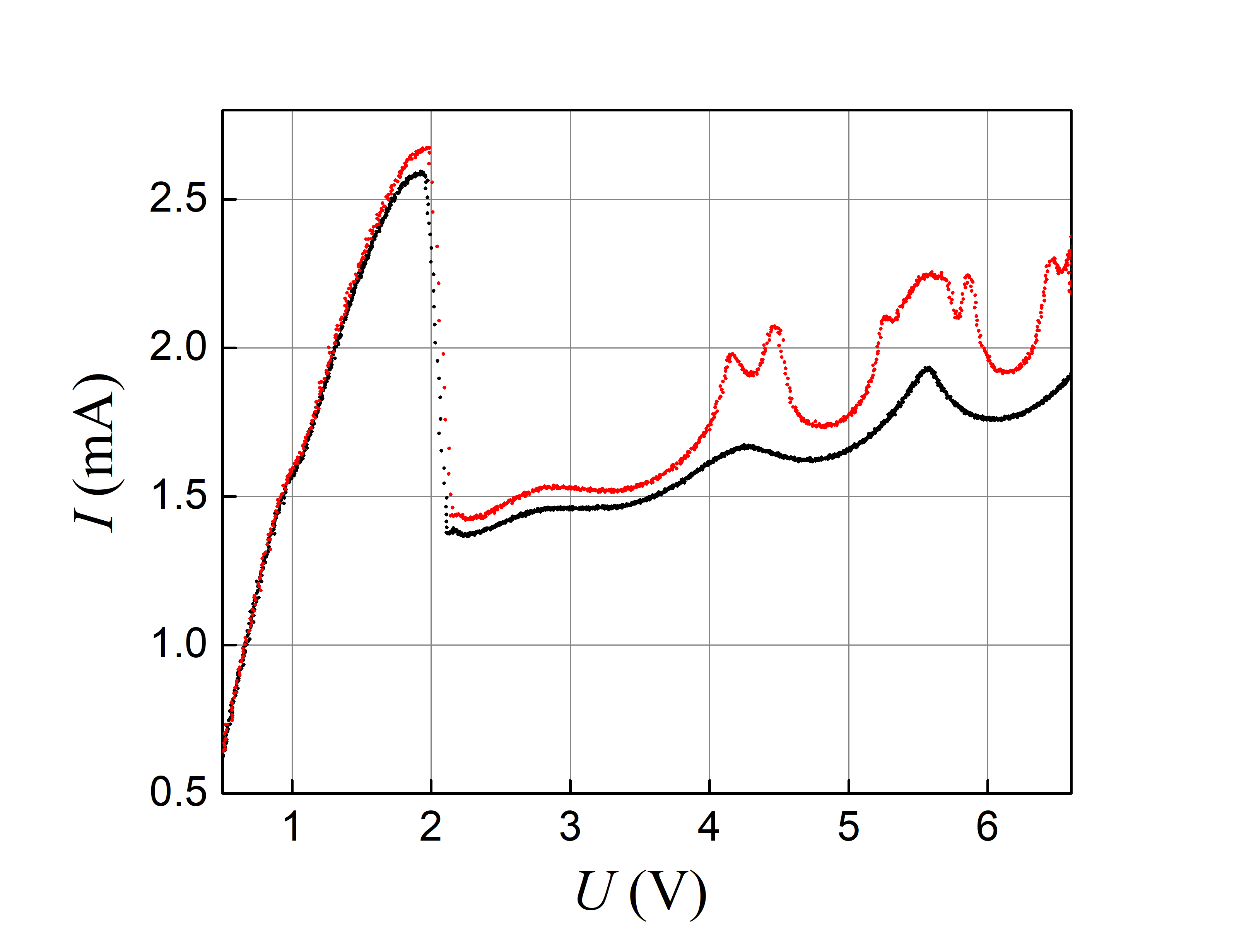}
    \caption{Current-voltage characteristics of GaAs/AlAs SL. Black -- in the dark, red -- under illumination.}
    \label{fig:fig3}
\end{figure}

\begin{figure}[h!] 
    \centering
    \includegraphics[scale=0.4]{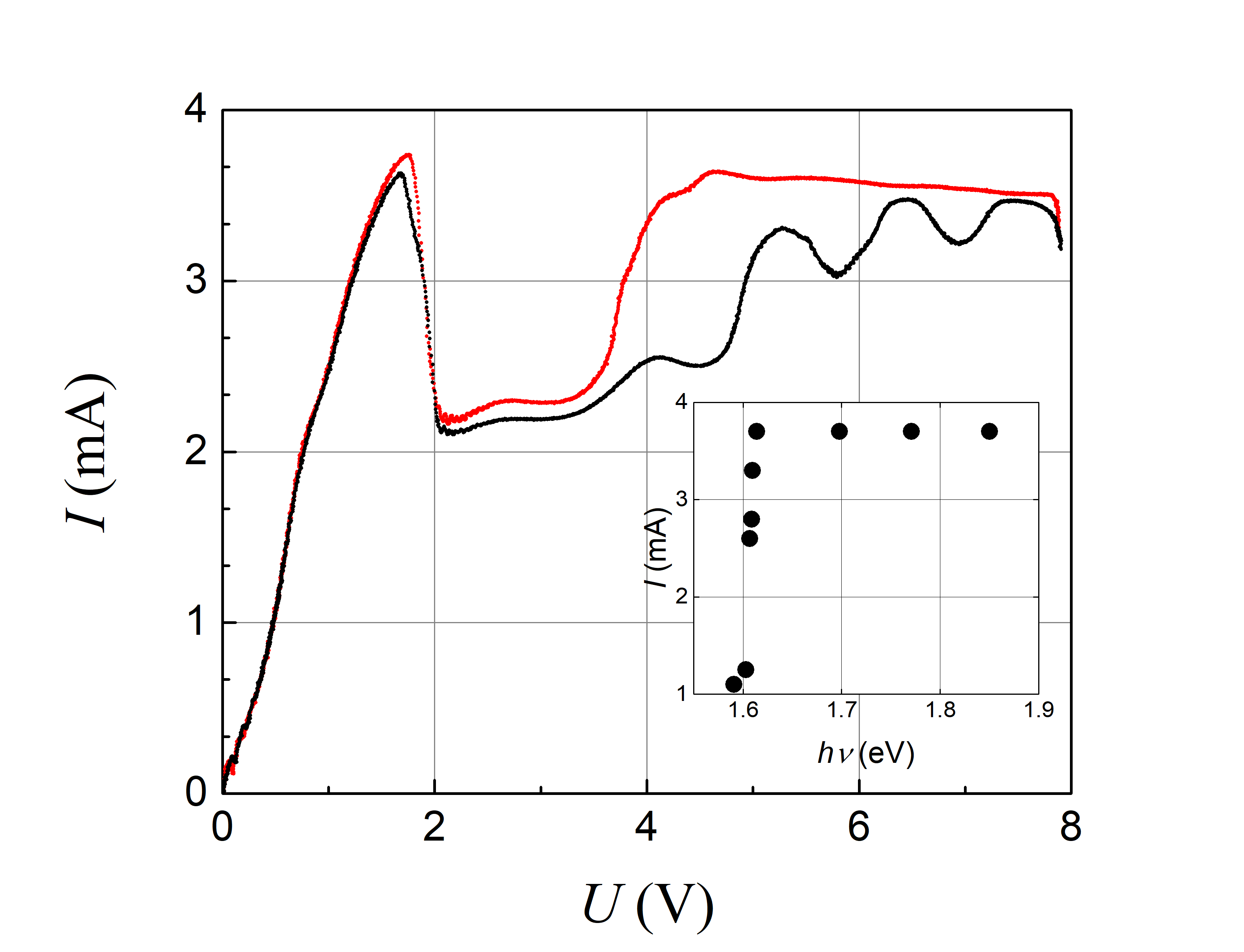}
    \caption{Current-voltage characteristics of GaAs/AlAs SL. Black -- in the dark, red  -- under illumination with intensity larger than in Figure~\ref{fig:fig2}. Inset: photoresponse spectrum.}
    \label{fig:fig4}
\end{figure}

\begin{figure}[h!] 
    \centering
    \includegraphics[scale=0.5]{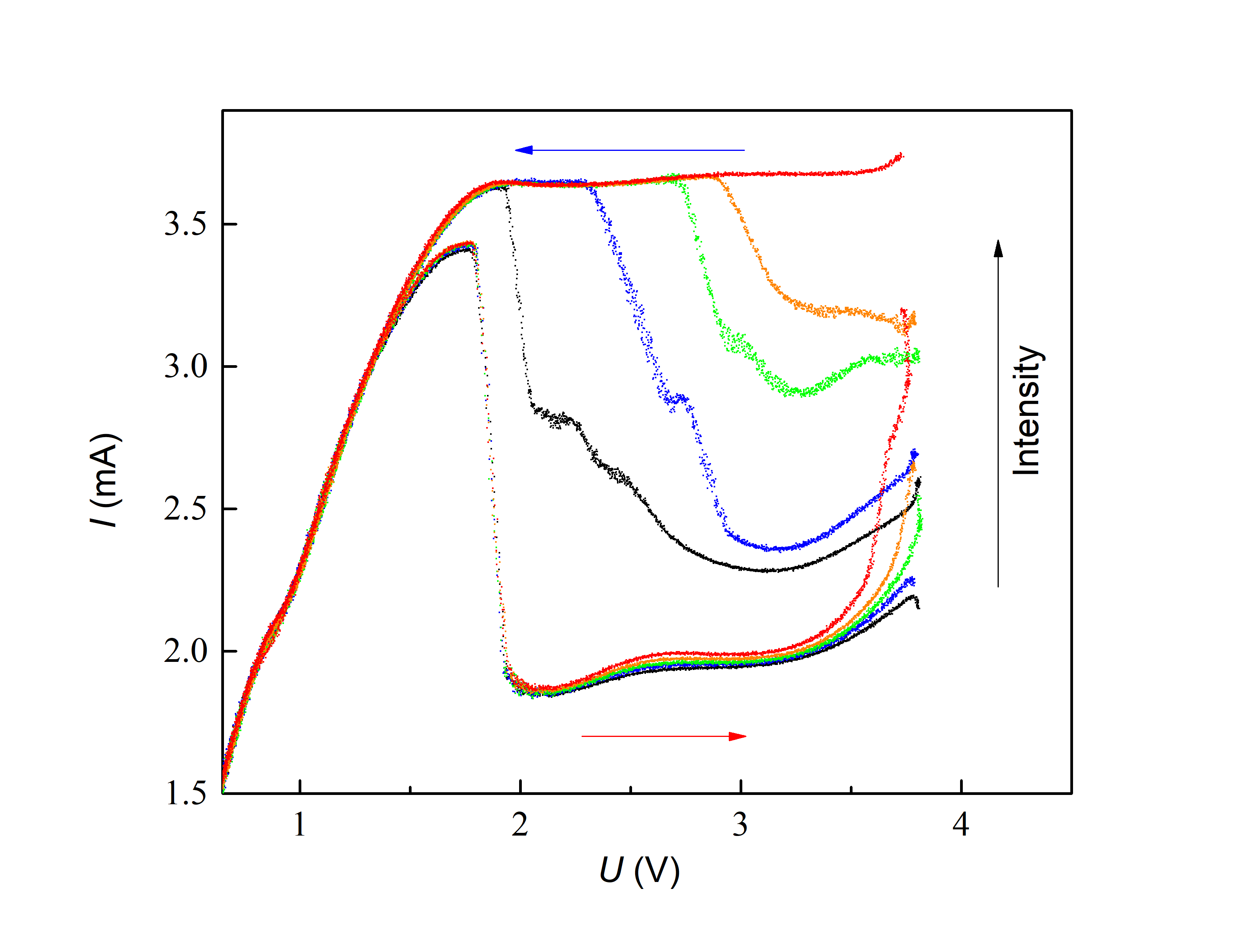}
    \caption{Current-voltage characteristics of GaAs/AlAs SL at different light intensities. 
Horizontal arrows show directions of voltage sweeps.
}
    \label{fig:fig5}
\end{figure}

A strong effect of a rather weak illumination (with a LED or $2.5$ W flashlight lamp) on a tunneling current in GaAs/AlAs SLs with relatively high doping level was found under conditions of electric domains formation. Figure~\ref{fig:fig3} shows the current-voltage characteristics with and without illumination. The same is shown in Figure~\ref{fig:fig4} at larger illumination intensity. 
Shown in Figure~\ref{fig:fig5} are parts of current-voltage characteristics at different light intensities. 
The increase in current under illumination could reach 50\% of the dark current. The sufficiently intense illumination could initiate the traveling-to-static domain transition, which becomes apparent in current saturation regions in I-V curves (Figure~\ref{fig:fig4}). The measured photocurrent spectrum (the inset in Figure~\ref{fig:fig4}) points out the band-to-band excitation. The spectral long-wavelength cutoff allows us to determine the GaAs QW gap and the energy of the first size quantization level. These energies coincide with known values for GaAs QW~\cite{bonilla}. It should be noted that the photocurrent at the static domain regime does not depend on exciting light frequency. 

At first sight, the large photoresponse turns out to be quite unexpected because the usual photoconductivity at room temperature has to be extremely weak. The reason is that the free electron concentration is of the order of the Si donors concentration ($\sim 10^{17}$\, cm$^{-3}$) as the donors with $\sim 6$\, meV binding energy have to be almost totally depopulated by thermal excitation. Indeed, at the voltage below threshold one (before the domain formation), the photoresponse did not observe. The high photoresponse was observed only at the domain formation regime. We attribute its origin to the triangular shape of the domain with the extended region of completely ionized donors. Because of the very small electron concentration inside the domain, the additional concentration of electron-hole pairs created by interband illumination changes the space charge at domain boundaries. This leads to a change of domain movement regime. The sufficiently intense illumination could initiate the traveling-to-static domain transition, which becomes apparent in the current saturation region in I-V curves (upper red curve in Figure~\ref{fig:fig4}). Thus, the high photoresponse of short-period GaAs/AlAs SLs at high voltages to the relatively weak interband illumination is connected with transformations of high-field domains rather than the rise of conductivity due to the increase of free electron concentration. The confirmation of the suggested explanation is also the independence of the photoresponse of the light spectrum when the illumination transforms the moving domain regime into the regime of the static domain. Indeed, the value of current in the saturation region at the static domain regime is determined only by the shape of the current-voltage characteristic of the initially uniform sample and does not depend on the way of entering this regime.

\subsection*{ACKNOWLEDGMENTS}
This work was carried out within the framework of the state task at Kotelnikov Institute of Radio Engineering and Electronics of RAS, and partially supported by RFBR projects 20-02-00624 and 20-52-16304. The equipment of the Unique Science Unit ''Cryointegral'' (USU \#352529) was used, which was supported by the Ministry of Science and Higher Education of Russia (Project No. 075-15-2021-667)

\bibliographystyle{unsrt}  


\end{document}